# What is missing from existing Lithium-Sulfur models to capture coin-cell behaviour?

Miss. Elizabeth Olisa, Dr. Monica Marinescu

10 March 2025

## Abstract

Lithium-sulfur (Li-S) batteries offer a promising alternative to current lithium-ion (Li-ion) batteries, with a high theoretical energy density, improved safety and high abundance, low cost of materials [1]. For Li-S to reach commercial application, it is essential to understand how the behaviour scales between cell formats; new material development is predominately completed at coin-cell level, whilst pouch-cells will be used for commercial applications. Differences such as reduced electrolyte-to-sulfur (E/S) ratios and increased geometric size at larger cell formats contribute to the behavioural differences, in terms of achievable capacity, cyclability and potential degradation mechanisms [2].

This work focuses on the steps required to capture and test coin-cell behaviour, building upon the existing models within the literature, which predominately focus on pouch-cells. The following areas have been investigated throughout this study, to improve the capability of the model in terms of scaling ability and causality of predictions.

- **Cathode surface area:** few models include detailed information about how much of the cathode surface area is available for reactions, based on the amount of electrolyte required to fill the cathode pores. Instead, the assumption is made that the entire area is available and all active material is fully utilised. Variations to physical cell parameters will impact this.
- **Precipitation dynamics:** most models assume only one precipitation mechanism occurs, which influences the behaviour consistently throughout cycling; the distinction is often not made between nucleation and particle growth precipitation mechanisms.
- **C-rate dependence:** capacity C-rate dependence is typically captured in higher order models through transport mechanisms. The inclusion of diffusion based on concentration gradients between regions in the cell can be utilised to capture C-rate variation within a low-order model.

## 1 Introduction

Lithium-sulfur (Li-S) batteries are a likely alternative chemistry to the current Li-ion batteries due to the significantly higher theoretical energy density [3], improved safety and the high abundance of low-toxicity sulfur [1]**.** Li-S utilises a conversion mechanism, as opposed to an intercalation mechanism like Li-ion; reactants form entirely new products [4], resulting in a change in the structure of the positive electrode. The electrode is made from a conductive carbon host, and sulfur, the insulating active material, is loaded on**.** Significant research is being carried out to understand variations in the electrode properties, and the most viable options for commercial application. Taking all areas into consideration, the material selection compromises battery performance in terms of highest capacity and cyclability, and the cost and scalability of the manufacturing processes [5].

For Li-S, new material development is predominately completed at coin-cell level due to the requirement for less material [2] and a simpler manufacturing process [6], [7], reducing the overall experimental time and cost. For commercial applications, pouch-cells are used due to the increased capacity density, cyclability and energy density [2]. There are numerous differences between these cell formats, discussed in detail for Li-ion cells by Bridgewater et al. [8]. The main physical differences contributing to behavioural differences that are relevant to Li-S cell performance are highlighted in Table 1. Whilst materials may perform highly when tested in coin-cells, this may not directly translate to larger cell formats. Understanding the reasons why, and the mechanisms that contribute to these differences is essential to reach commercial application.

| Component | Coin-Cell | Pouch-Cell |
| --- | --- | --- |
| Cathode | Small, circular, single-sided | Large, rectangular, double-sided |
| Layers | Single | Multiple |
| Weight | Excess electrolyte contributing to 'dead' weight (low energy density) | Less excess of materials (high energy density) |
| Packaging | Consistent, uniform internal pressure (hard casing, spacers, spring) | Variable, external pressure (soft casing) |
| Electrolyte to Sulfur (E/S) Ratio | High: large electrolyte excess ensuring full cathode utilisation, and preventing electrolyte as limiting factor | Low: minimal electrolyte to achieve high energy density |

*Table 1: Key Physical Differences between Coin and Pouch-Cells. A more detailed table is included in [2].*

The development of a trustworthy model can accelerate the experimental testing process. For example, running simulations to determine optimum cell parameters would be significantly faster than designing, building and testing various cells. Where there are multiple hypotheses for observed behaviour, a model becomes a useful tool to test these efficiently, determining which is most likely, or contributes most significantly, and mitigate the least likely. Where model results do not align, or disprove the hypotheses, further work can be undertaken to determine the reasoning and investigate alternative mechanisms contributing to the behaviour.

Physics-based modelling aims to analyse what is happening inside a cell by focusing on the mathematics defining these mechanisms. There are many highly regarded published physics-based models within the literature. One of the first studies by Mikhaylik and Akridge [9] investigated the influence of polysulfide shuttle, capturing charge and discharge in a one-dimensional (1D) model. Kumaresan et al. [10] developed a general 1D model, based on the 1D model by White et al. [11], with the inclusion of variable porosity and area. Zhang et al. [12] improved the prediction for resistance in a zero-dimensional (0D) model. Zhang et al. [13] then developed another 1D model to capture discharge at high C-rates based on the transport mechanisms. Marinescu et al. [14] built upon existing 0D models to capture charge and discharge, along with investigating the influence of the precipitation mechanism. Further work by Marinescu et al. [15] provided a model to capture reversible and irreversible capacity change during cycling. Following on from earlier work, Zhang et al. [16] provided another 0D model investigating the influences on rate capability within Li-S. Cornish et al. [17] provided a detailed review of the existing models and their capability, contributing towards an improved 0D model.

The study by Cornish et al. [17] showed that whilst the fundamental features required to model cell behaviour were collectively included in this published work, a single model rarely captured all experimentally observed features. For example, of the 24 models investigated, only 6 captured cycling voltage through both charge and discharge mechanisms. With the aim of a well-developed coin-cell model to be used alongside experimental work to test new materials, it is essential a model includes these key features. Alongside this, many of the published models are not available open source, making it difficult to test their validity beyond the scope of the original paper. Those that are available and capture the required experimental features are designed and parameterised for pouch-cells; to the best of our knowledge, there are no physics-based, cell-level coin-cell models available within the literature. Due to the physical and behavioural differences between cell formats, without vigorous testing it is unclear how well these pouch-cell models can be used to predict coin-cell behaviour.

Therefore, the aim of this study was to investigate the capability of existing pouch-cells models to predict coin-cell behaviour, and highlight any additional features required to improve these predictions. This works towards confirming which mechanisms retrieve the results observed in different cell-formats, improving the understanding of the scale-up process from coin-cell development to pouch-cell application. With the philosophy of starting with the simplest method, this study focused and built upon 0D models, due to low computational cost and requirement for minimal fitting parameters and assumptions. Three main areas were focused upon: cathode surface area, the precipitation dynamics, and C-rate dependence. This study showed the impact of updating and incorporating additional features to capture both coin and pouch-cell behaviour, through changing the required scaling parameters to match experimental results. The selection of experimental data for parameterisation, along with additional model requirements is discussed in Section 2: Considerations. An overview of the model formulation is discussed in Section 3: Model Development. A summary of the impact of all the model upgrades is discussed in Section 4: Model Results. The reasoning for these changes, the key findings and further work to improve the capability of the model is discussed in Appendix, Section 7: *Model Iterations and Discussion*.

## 2 Considerations (Capturing Coin-Cell Experimental Behaviour)

### 2.1 Parameterisation: Boenke et al. [18]

Several experimental papers were reviewed to use to parameterise the model, including work by Wang et al. [19], Chen et al. [20] and Weller et al. [21], who all investigated the influence of the electrolyte on performance. Schmidt et al. [22] investigated the influence of swelling of cathodes produced by dry processing, and Boenke et al. [18] investigated scalable cathode manufacturing methods. After review, data by Boenke et al. [18] was selected to parameterise the model, as it included detailed information and results for both coin and pouch-cells, built and tested under the same condition, enabling comparisons as shown in Figure 1. The study also included detailed information regarding the manufacturing and characterisation of materials, reducing the number of assumptions the model would require. The materials and techniques used were common within the field, meaning when using this data set for parameterisation, the model predictions would likely be applicable to a large range of other experimental work.

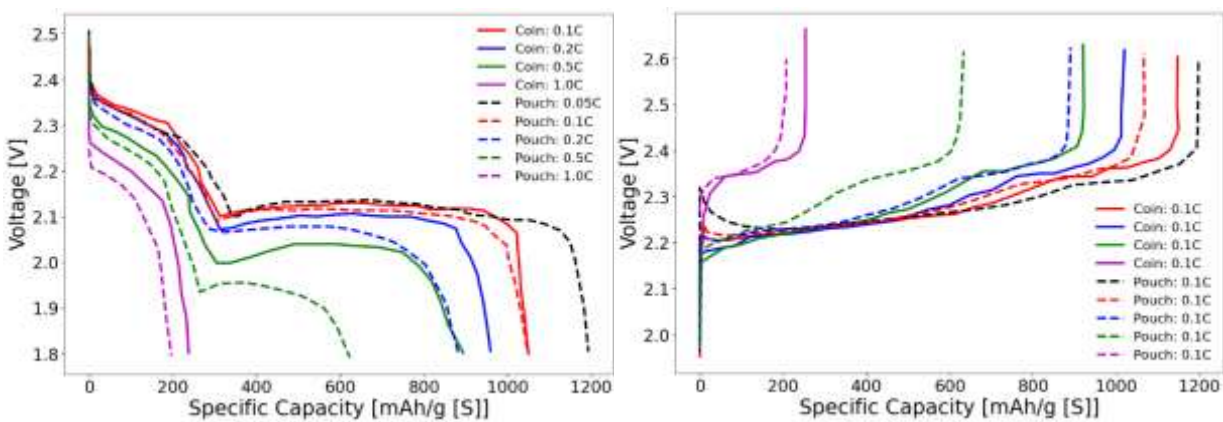

*Figure 1: Boenke et al. [18] Coin and Pouch Discharge (left) and Charge (right) Experimental Results, for CNT BP Cathode*

### 2.2 Specific Model Requirements (Mechanisms)

To test the capability of a model, it is essential to understand which mechanisms cause the observed behaviour, and the extent to which these may differ between a coin and pouch-cell. The features from existing models that capture cycling behaviour are summarised in Table 2. The highlighted behaviour was focused on throughout this study.

| Observed Behaviour | Model Feature | Achieved in 0D? |
|---|---|---|
| Cycling voltage plateaus | (Reversible) charge and discharge reactions (with varying number of species to capture correct plateaus) Precipitation dynamics, relating to voltage kinks and jumps | ✓ |
| Variable area | Pore blocking due to precipitation | ✓ |
| (Irreversible) Capacity change | Degradation (e.g. loss of active material, polysulfide shuttle) | ✓ |
| (Reversible) Capacity change | Transport mechanisms and limitations | ✗ |
| C-rate voltage dependence | Resistance predictions (and inclusion of ohmic change) | ✓ |
| Performance variation with cell format | Model inputs (e.g. area, mass E/S ratio, layers, casing), relating to reactions, material utilisation, transport etc. | ✗ |

*Table 2: A summary of the observed behaviour, and the features within a model which capture this behaviour.*

### 2.2.1 Charge and Discharge

One of the fundamental requirements to make comparisons to the experimental results is the inclusion of both the charge and discharge mechanisms within the model. This is achieved through reversible electrochemical and chemical reactions. It is common for models to focus on only one direction of reaction; this is hypothesised to be due to the complexity of the varying mechanisms that occur during different directions of reaction, including polysulfide shuttle and dissolution of species in charge [10], and precipitation and pore blocking during discharge [23]. As shown in the review by Cornish et al. [17], 19 of the 26 models analysed captured discharge, 11 captured charge and only 6 captured both.

The number of species involved in the charging and discharging reactions varies between models. Typically, the number is selected based on what the model is aiming to achieve. Further experimental work is required to confirm the exact number present, and under which conditions. For the purpose of this model, a sufficient number of species were required to capture the charge and discharge plateaus observed in the experimental data, and the precipitation and dissolution mechanisms. An ideal model should include the capability to predict and test the influence of degradation mechanisms, including polysulfide shuttle.

### 2.2.2 Area Changes

Variations to the cathode surface area are captured in existing models through the change in porosity, as derived by Kumaresan et al. [10]. As this model only captures discharge, the implementation by Zhang et al. [12] was studied in detail. Whilst variations to the fitting parameter in the equation can capture varying magnitudes of area change, this equation does not account for the influence of precipitation on the available surface area directly. Based on the derivation of the porosity equation in these models [10] [12], precipitation influences porosity, and the magnitude of porosity change in relation to the initial porosity determines the available area. Under certain conditions, this ratio term could result in a much smaller prediction of the variation in area compared to experimentally observed changes.

### 2.2.3 Capacity Change

A varying discharge current is required to test C-rate dependency, whilst the charging C-rate remained constant in the experimental study by Boenke et al. [18]. It was found that the capacity varied with C-rate in both discharge and charge [18], hypothesised to be due to the history effect. Long-term cycling under a constant discharge C-rate was also carried out, enabling comparisons when investigating rate dependence and capacity fade.

Higher order models capture reversible capacity C-rate dependence through the inclusion of transport mechanisms and limitations [13]. Low order models have attempted to account for transport and the limitations through a modified Butler-Volmer equation [1], based on limiting the current. Separately, irreversible capacity fade is captured through degradation mechanisms such as polysulfide shuttle [9], [24], or loss of active material [23]. Degradation mechanisms are also influenced by C-rate; a larger capacity loss at higher C-rates can occur due to uneven plating on the lithium anode [25], or a reduced influence of polysulfide shuttle due to higher competing rates of reaction [23].

A study by Brückner et al. [26] investigated the influence of manufacturing and cycling variations on cycle stability and sulfur utilisation and found that capacity fade reduced at higher C-rates, higher electrolyte volumes and lower sulfur loadings. Fan et al. [27] investigated the influence of C-rate whilst focusing on precipitation mechanisms and found that capacity decreased with higher C-rate, relating to the type and morphology of precipitates. Mechanisms to capture both nucleation and particle growth precipitation have only been captured in several 1D models [28], [29].

### 2.2.4 Voltage Variation

The voltage C-rate dependence is captured through the Ohmic change, based on the resistance calculation derived by Zhang et al. [12]. In pouch-cells, alongside a higher resistance, a higher current flowing through the cell results in a larger change in the cell voltage due to the Ohmic change, compared to coin-cells [2]. A voltage kink can be observed at the transition from high to low plateau during discharge, caused by an accumulation of $S_1$ before it begins precipitating [15]. A voltage jump at the beginning of charge corresponds to a dissolution bottleneck, caused by slow dissolution of precipitated species, reducing the availability of species for further reactions [15]. Through parameter selection and formulation of the precipitation dynamics, the model should be able to capture the impact of variations in resistance and voltage between cell formats.

### 2.2.5 Scaling between Cell Formats

To capture variations between coin and pouch-cells, the model needs to include scaling parameters, including the electrolyte-to-sulfur (E/S) ratio, the mass of sulfur, the geometric area, the number of layers and the current through the cell. Due to the limited number of models available open source, it was unclear how these parameters were included, calculated and updated in existing models. The mechanisms relating to these parameters also differ; for example, coin-cells have a larger excess of electrolyte, reducing the concentration of species, electrolyte viscosity and resistance within the cell [23]. The change in quantities may also influence some of the degradation mechanisms; for example, as pouch-cells have an increased geometric size and areal current density, polysulfide shuttle has been found to occur more readily, alongside rapid dendrite growth [30].

## 3 Model Development

The final version of this model has been implemented in PyBaMM [31], and is available with this publication. The nomenclature is summarised in Table 3 in the Section 7.1. The model was developed based on existing models, as discussed throughout the study, combining the key features required to capture observed behaviour. The base model used was by Cornish et al. [17], due to the detailed testing and validation, and availability of the model open source.

To capture the cycling behaviour, and the ratio of high to low plateau capacity as observed by Boenke et al. [18], three sulfur species were selected: $S_8$, $S_4$ and $S_1$. The reactions to form each species are shown in Equations (1) and (2). Only $S_1$ precipitates at the end of discharge.

$$S_8^0 + 4e^- \leftrightarrow 2S_4^{2-} \tag{1}$$

$$S_4^{2-} + 6e^- \leftrightarrow 4S_1^{2-} \downarrow \tag{2}$$

The first model upgrade required to capture the observed behaviour involved making a clear differentiation between the porosity, ε, and the pore volume, $v_{pore\ (cath)}$. This enabled a more accurate calculation of the usable electrolyte volume, v. The pore volume was calculated as shown in Equations (5) - (8), based on the volume of the cathode structure and volume of sulfur. The usable electrolyte volume was calculated as shown in Equations (4), (9), (10), (12), (13), and describes the volume of electrolyte contained within the cathode pores, therefore contributing to electrochemical reactions. This volume was used when calculating the concentration of each species, as shown by Equation (3). It is important to note the concentration of precipitated sulfur was included for consistency but is not a physically meaningful parameter when calculated this way, as solid precipitates are no longer dissolved within the electrolyte.

The usable electrolyte volume was also used to calculate the initial available surface area, as shown in Equations (10) – (11)(14). The electrolyte volume, along with the sulfur species dissolved within, that was not contained within the cathode pores and therefore not considered usable for electrochemical reactions, was considered as excess. This is discussed in detail in Section 7.2.

$$C_{Sx} = \frac{S_x}{Ms \cdot ns_x \cdot v} \tag{3}$$

$$v_{elec} = \frac{E}{S} \cdot S_{tot} \tag{4}$$

$$v_{cath(CNT)} = \frac{cath_{mass}}{\rho_{single_{CNT}}} = \frac{\rho_{cnt_{pure}} \cdot (ar_{geom} \cdot x_{cath}) \cdot l}{\rho_{single_{CNT}}} \tag{5}$$

$$v_{sulfur} = \frac{S_{total}}{\rho_{sulfur}} \tag{6}$$

$$v_{tot} = (ar_{geom} \cdot x_{cath}) \cdot l \tag{7}$$

$$v_{pore\ (cath)} = v_{tot} - v_{cath(CNT)} - v_{sulfur} \tag{8}$$

$$\text{If } v_{pore} < v_{elec}: \tag{9}$$

$$v = v_{pore\ (cath)}, \tag{10}$$

$$\tag{11}$$

$$ar_0 = ar_{spec} \cdot (cath_{mass} - S_{tot})$$

$$\text{If } v_{pore\ (cath)} \geq v_{elec}: \tag{12}$$

$$v = v_{elec}, \tag{13}$$

$$ar_0 = ar_{spec} \cdot (cath_{mass} - S_{tot}) \cdot \left(\frac{v}{v_{pore\ (cath)}}\right) \tag{14}$$

Alongside introducing the pore volume term, the porosity equation was also updated, as shown in Equations (15) and (16). The new equation accounted for the influence of precipitation leading to pore blocking, and was calculated based on the total cathode volume, the volume of the CNT structure of the cathode and the volume of precipitated sulfur.

$$v_{S_p} = \frac{S_p}{\rho_{sulfur}} = \frac{C_{Sp} \cdot Ms \cdot v \cdot ns1}{\rho_{sulfur}} \tag{15}$$

$$\varepsilon = \left(1 - \frac{V_{cath\,(CNT)}}{V_{tot}} - \frac{V_{S_p}}{V_{tot}}\right) \tag{16}$$

The second model upgrade, in addition to pore volume and porosity, involved recalculating area to account for the influence of precipitation by particle growth and nucleation mechanisms separately, as shown in Equations (17) and (18). Only precipitation via nucleation influenced the area change, through the formation of nucleation sites. This removed the existing assumption that precipitation of sulfur influenced the area consistently throughout cycling. An additional growth area term, $ar_{growth}$, was introduced to account for the area formed due to particle growth precipitation, as shown in Equation (19). These changes enabled a larger magnitude of area change due to precipitation, and a dependence of area change on C-rate. This is discussed in Section 7.3.

$$\gamma = \frac{\pi \cdot r^2 \cdot A}{Ms} \div x_{nuc} \tag{17}$$

$$ar = ar_0 - (\gamma \cdot Sp_{(nuc)}) \tag{18}$$

$$ar_{growth} = (3 \cdot (C_{Sp_{(growth)}} \cdot v \cdot VLi_2S))^{\frac{2}{3}} \tag{19}$$

To capture voltage C-rate dependence, the model accounted for the Ohmic change, calculated using the resistance equation derived by Zhang et al. [12], as shown in Equations (21) and (22). In this model, the concentration of lithium was calculated in relation to the concentration of charged sulfur species, $S_1$ and $S_4$, as shown in Equation (20).

$$C_{Li} = 2 \cdot (C_{S1} + C_{S4})) \tag{20}$$

$$conduc = \varepsilon^z \cdot \left(conduc_0 - (b \cdot |C_{Li} - C_{Li_0}|)\right) \tag{21}$$

$$res = \frac{l}{ar_{geom} \cdot conduc} \tag{22}$$

The reaction shown in Equation (1) was referred to as the high plateau reaction, and Equation (2) as the low plateau reaction. The rates of each reaction were described by the Butler Volmer equations, (23) and (24) respectively. These used the fixed standard potentials rather than variable Nernst potentials, allowing the simplest model formulation to ensure stability when running under varying conditions, without numerical errors occurring. The partial currents, $i_L$ and $i_H$, for each reaction sum together to make the total applied current, as shown in Equation (25). The cell voltage was calculated from the terminal voltage and the Ohmic change, as shown in Equation (26).

$$i_H = (-i_H^0 \cdot ar) \cdot \left(C_{S4}^2 \cdot e^{\frac{n_H F}{2RT}(V-E_H^0)}\right) - \left(\frac{C_{S8}}{e^{\frac{n_H F}{2RT}(V-E_H^0)}}\right) \tag{23}$$

$$i_L = (-i_L^0 \cdot ar) \cdot \left(C_{S1}^4 \cdot e^{\frac{n_L F}{2RT}(V-E_L^0)}\right) - \left(\frac{C_{S4}}{e^{\frac{n_L F}{2RT}(V-E_L^0)}}\right) \tag{24}$$

$$I = i_H + i_L \tag{25}$$

$$V = V_{terminal} - (I \cdot res) \tag{26}$$

The movement of species during cycling was governed by a differential system of equations accounting for the electrochemical and chemical reactions. The final model upgrade involved the addition of a diffusion mechanism between species in each of the four regions within the cell: cathode, separator, cathode reservoir and separator reservoir. During cycling, species could be regained from other regions into the cathode through this diffusion mechanism. This was dictated by the magnitude of the transport coefficients. This is discussed in detail in Section 7.4. With the inclusion of diffusion, this model should now be considered as a pseudo-spatial 0D model.

The reactions that occur within the cathode are shown in Equations (27) - (31). The cathode reservoir reactions are shown in Equation (32), the separator reservoir in Equation (33), and the separator in Equation (34), where x corresponds to the sulfur species: $S_8$, $S_4$ and $S_1$. The separator reactions for $S_8$ and $S_4$ contain the shuttle reaction ($k_s \cdot C_{S8}$) and ($2 \cdot k_s \cdot C_{S8}$) respectively. For the results shown throughout the study, the shuttle rate was set to zero.

$$\frac{dC_{S_8^0}}{dt} = -\left(\left(\frac{1}{nH}\right) \cdot \left(\frac{i_H}{F \cdot v}\right)\right) - \left(\left(D_{cath_{res_{S8}}} \cdot \left(C_{S8} - C_{cath_{res_{S8}}}\right)\right) \cdot \frac{v_{cath_{res}}}{v + v_{cath_{res}}}\right) \quad (27)$$
$$- \left(\left(D_{S8} \cdot \left(C_{S8} - C_{sep_{S8}}\right)\right) \cdot \frac{v_{sep}}{v + v_{sep}}\right)$$

$$\frac{dC_{S_4^{2-}}}{dt} = \left(\left(\frac{2}{nH}\right) \cdot \left(\frac{i_H}{F \cdot v}\right)\right) - \left(\left(\frac{1}{nL}\right) \cdot \left(\frac{i_L}{F \cdot v}\right)\right) \quad (28)$$
$$- \left(\left(D_{cath_{res_{S4}}} \cdot \left(C_{S4} - C_{cath_{res_{S4}}}\right)\right) \cdot \frac{v_{cath_{res}}}{v + v_{cath_{res}}}\right)$$
$$- \left(\left(D_{S4} \cdot \left(C_{S4} - C_{sep_{S4}}\right)\right) \cdot \frac{v_{sep}}{v + v_{sep}}\right)$$

$$\frac{dC_{S_1^{2-}}}{dt} = \left(\frac{4}{nL}\right) \cdot \frac{i_L}{F \cdot v} - \left(\left(D_{cath_{res_{S1}}} \cdot \left(C_{S1} - C_{cath_{res_{S1}}}\right)\right) \cdot \frac{v_{cath_{res}}}{v + v_{cath_{res}}}\right) \quad (29)$$
$$- \left(\left(D_{S1} \cdot \left(C_{S1} - C_{sep_{S1}}\right)\right) \cdot \frac{v_{sep}}{v + v_{sep}}\right)$$
$$- \left(ar \cdot C_{S1} \cdot k_{nuc} \cdot C_{Sp_{nuc}} \cdot e^{-F \cdot \frac{V-E_{nuc}}{RT}}\right)$$
$$- \left(\frac{Ms}{\rho} \cdot \left(ar_{growth} + (ar_0 - ar)\right) \cdot k_{p/d} \cdot (CS^{2-} - CS^*) \cdot C_{Sp_{growth}}\right)$$

$$\frac{dC_{Sp_{nuc}}}{dt} = ar \cdot C_{S1} \cdot k_{nuc} \cdot C_{Sp_{nuc}} \cdot e^{-F \cdot \frac{V-E_{nuc}}{RT}} \quad (30)$$

$$\frac{dC_{Sp_{growth}}}{dt} = \frac{Ms}{\rho} \cdot \left(ar_{growth} + (ar_0 - ar)\right) \cdot k_{p/d} \cdot (CS^{2-} - CS^*) \cdot C_{Sp_{growth}} \quad (31)$$

$$\frac{dC_{cath_{res_{Sx}}}}{dt} = -\left(\left(D_{res_{Sx}} \cdot \left(C_{cath_{res_{Sx}}} - C_{sep_{res_{Sx}}}\right)\right) \cdot \frac{v_{sep_{res}}}{v_{cath_{res}} + v_{sep_{res}}}\right) \quad (32)$$
$$- \left(\left(D_{cath_{res_{Sx}}} \cdot \left(C_{cath_{res_{Sx}}} - C_{Sx}\right)\right) \cdot \frac{v}{v + v_{cath_{res}}}\right)$$

$$\frac{dC_{sep_{res_{Sx}}}}{dt} = \left(\left(D_{res_{Sx}} \cdot \left(C_{cath_{res_{Sx}}} - C_{sep_{res_{Sx}}}\right)\right) \cdot \frac{v_{cath_{res}}}{v_{cath_{res}} + v_{sep_{res}}}\right) \quad (33)$$
$$- \left(\left(D_{sep_{res_{Sx}}} \cdot \left(C_{sep_{res_{Sx}}} - C_{sep_{Sx}}\right)\right) \cdot \frac{v_{sep}}{v + v_{sep}}\right)$$

$$\frac{dC_{sep_{Sx}}}{dt} = \left( \left( D_{sep_{res_{Sx}}} \cdot \left( C_{sep_{res_{Sx}}} - C_{sep_{Sx}} \right) \right) \cdot \frac{v_{sep_{res}}}{v_{sep_{res}} + v_{sep}} \right) \quad (34)$$
$$- \left( \left( D_{Sx} \cdot \left( C_{sep_{Sx}} - C_{Sx} \right) \right) \cdot \frac{v}{v + v_{sep}} \right)$$

## 4 Model Results

The model predictions when all three upgrades are accounted for, and then scaled between coin and pouch-cell are shown in Figure 2. The results at all C-rates tested by Boenke et al. [18] are included in the Appendix (Section 7.5). The aim of the model, as discussed in Section 2.2 was to capture scaling variations between cell formats, along with improving predictions for the magnitude of area change and capture C-rate dependence in a 0D model. To do this, the charge and discharge behaviour and the corresponding mechanisms were required, but degradation was not accounted for.

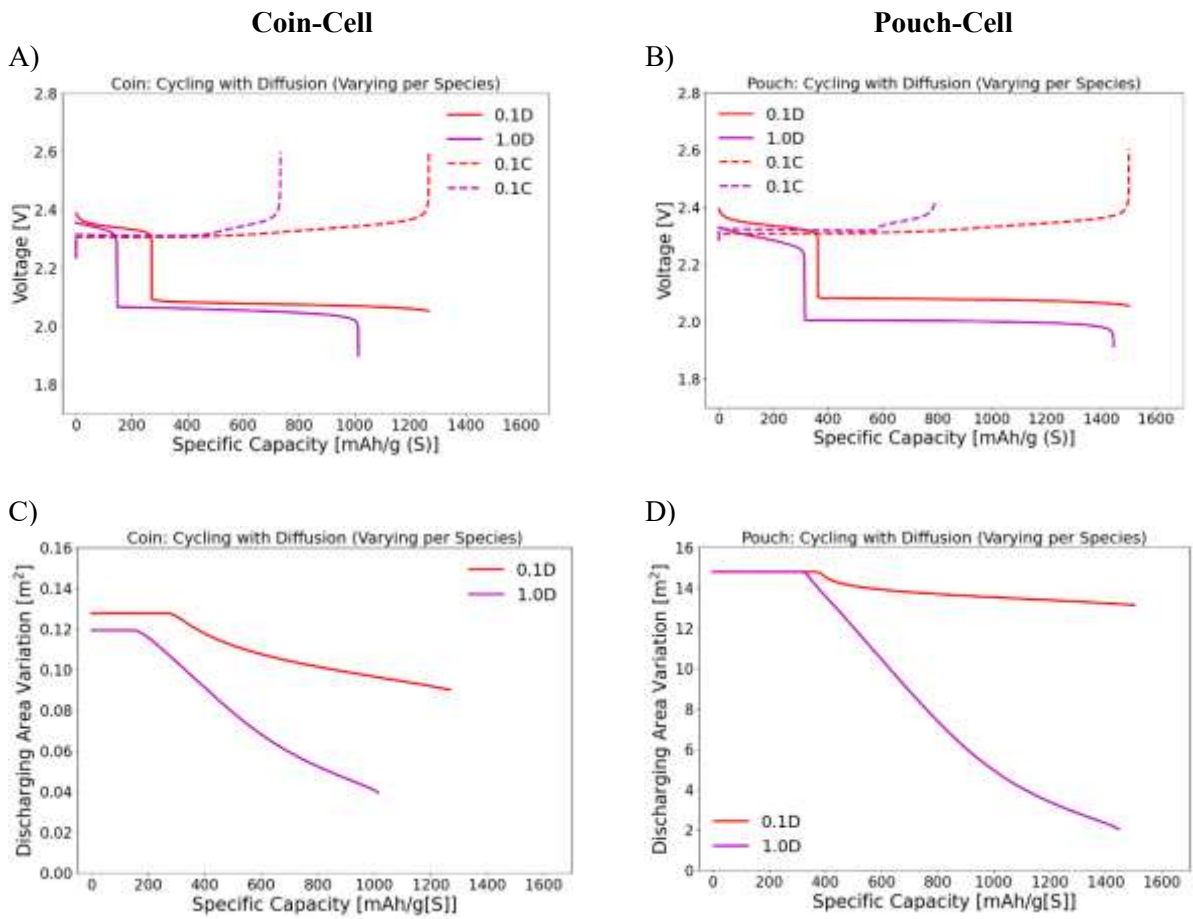

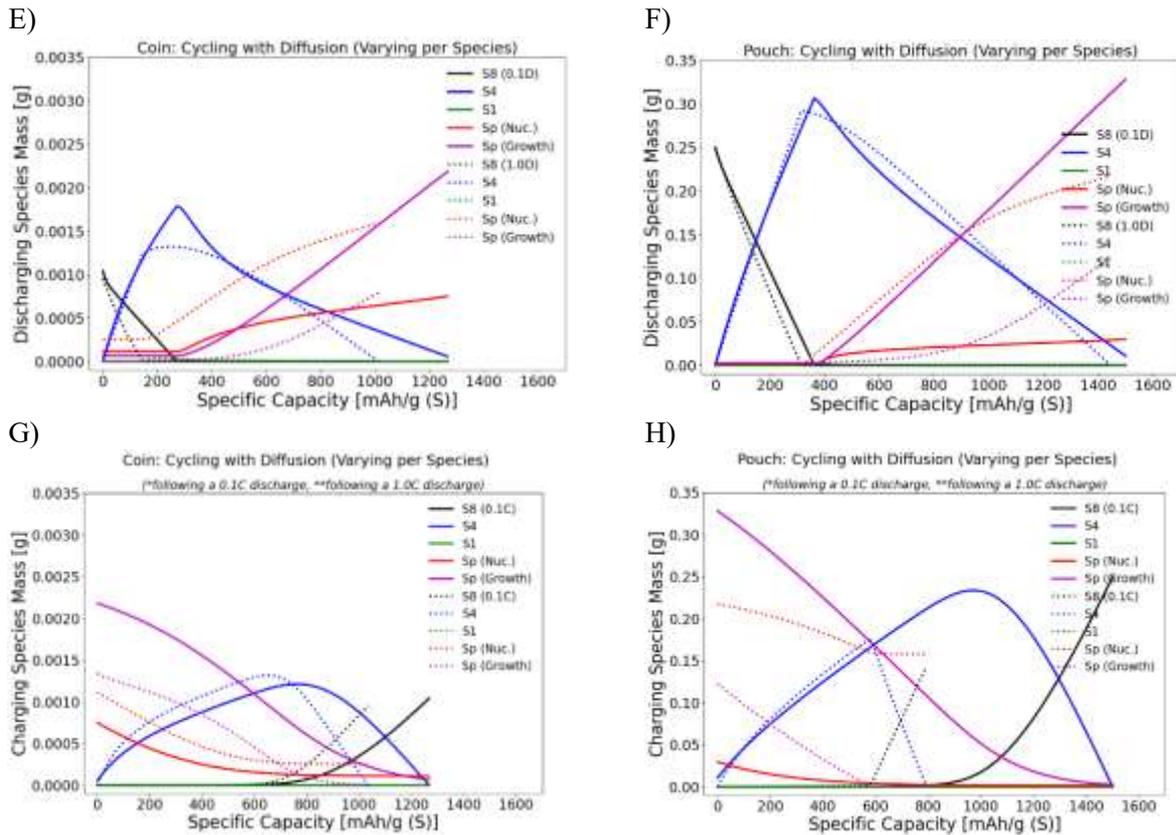

*Figure 2: Model predictions for coin-cell vs. pouch-cell behaviour.*

### 4.1 C-rate Dependence

With all three upgrades, the model predicts both charge and discharge capacity C-rate dependence, in both cell formats, as shown in Figure 2A and B. This was captured by the distribution of active material throughout the cell. The sulfur species were assumed to be homogeneously distributed throughout the electrolyte, which was contained across the four regions: cathode, separator, cathode reservoir and separator reservoir. Only species dissolved within the electrolyte contained in the cathode region contributed to electrochemical reactions. As the cell cycled, species diffused between these regions based on the concentration gradients that formed. At higher C-rates, if the higher rates of reaction were competing against lower rates of diffusion, bottlenecks began to form, preventing species being transported into the cathode to contribute towards electrochemical reactions. This resulted in lower achievable capacity at higher currents, resulting in the observed C-rate variation.

The capacity C-rate dependence was more significant in coin-cells than pouch-cells; a larger volume of excess electrolyte in coin-cells meant a larger proportion of species not contributing to reactions, and potentially larger diffusion bottlenecks, reducing the achievable capacity. This differed to the experimental results observed by Boenke et al. [18] in Figure 1, where coin-cells achieved higher capacity in both charge and discharge in comparison to pouch-cells. One hypothesis is that the model only includes the influence of diffusion, whilst experimentally, as pouch-cells have a smaller excess of electrolyte, the species concentration and electrolyte resistance would be higher, along with increased viscosity [18], which would increase the transport limitations and influence the C-rate dependence [13]. In the model, the smaller excess of electrolyte in the pouch-cell resulted in a larger proportion of the sulfur being contained within the cathode, rather than the reservoir. As more species were available for electrochemical reactions, bottlenecks relating to diffusion were reduced, and a higher capacity was achieved. As viscosity and other transport mechanisms were not accounted for in this model, the reduced impact of diffusion outweighed the increased resistance. This was further tested by varying the E/S ratio within the cell, which confirmed the same result; a larger C-rate dependence was observed with a larger excess of electrolyte.

### 4.2 Cathode Surface Area Variation

To improve the causality of predictions against experimental results, the model included two separate precipitation mechanisms. During discharge, the nucleation mechanism occurred first and was driven by voltage, followed by particle growth, which was driven by saturation concentration. This is discussed in detail in Section 7.3.

Only precipitation via nucleation influenced the cathode surface area; particle growth occurred on top of existing nucleation sites, resulting in minimal influence. The model prediction for the change in area during cycling followed the behaviour of nucleation, as shown in Figure 2C, E and D, F; there was no change in high plateau, as expected, and then the area began to decrease as nucleation began at the transition from high to low plateau in discharge. As more area was covered, the rate of nucleation decreased, as shown by Equation (30), allowing the rate of particle growth to increase, as shown by Equation (31), causing the change in area to tend towards a plateau. This change was dictated by the magnitude of the driving forces for each precipitation reaction.

For coin-cells, under the cycling conditions selected, the model predicted incomplete dissolution of nucleated species during charge, as shown in Figure 2G, resulting in a temporary loss in active material. As seen in Figure 2C, this incomplete dissolution also resulted in a loss in active surface area, shown by the maximum surface area at the beginning of discharge decreasing with increasing C-rate. It was hypothesised that at higher currents, there was less time for the dissolution reaction, leaving a larger proportion of species in precipitated form, blocking the pores and reducing the available area for electrochemical reactions to take place. The reduction in both species and area available for reactions contributed towards the capacity change observed. This was confirmed when further cycles at a lower C-rates were run, and the quantity of species dissolving during charge increased, increasing the species and area available for reactions.

Within the model, the magnitude of area change and the corresponding behaviour varied dependent on the precipitation, dissolution and nucleation rates, all of which were fitting parameters. For example, a voltage kink at the transition between plateaus in discharge could be observed when the relative rates of reaction were high in comparison to the precipitation rate, allowing $S_1$ to accumulate before precipitating.

### 4.3 Scaling between Formats

The inclusion of the scaling parameters such as the E/S ratio, geometric size and cycling current allowed the model to easily test the variations in coin and pouch-cell behaviour. For example, the direct inclusion of the E/S ratio ensured the concentration of species was correctly calculated under all conditions. As the current used in the model were calculated from the experimental current density for coin-cells given by Boenke et al. [18], the inclusion of the cell geometry enabled the current to accurately be calculated when scaling between cell formats. Aligning with experimental results as shown by Boenke et al. [18], the Ohmic change was larger in pouch-cells than coin-cells, due to the increased current and increased resistance. The voltage C-rate dependence was captured similarly to existing models, through this inclusion of Ohmic changes [12]. As the C-rate remained constant throughout each charge cycle to match experimental results by Boenke et al. [18], no voltage change occurred.

### 4.4 Boenke et al. Experimental Hypotheses

Beyond the areas previously discussed, Boenke et al. [18] made several hypotheses for the observed behaviour; this section discusses the capability of the model to predict these features.

Boenke et al. [18] found that the achievable capacity of the cell increased with reduced sulfur weight percent (wt%), as a lower mass resulted in more area available for reactions. It was also found that a lower sulfur loading increased capacity due to the reduced internal resistance [18]. The model captured an increased initial capacity and improved rate capability with reduced sulfur loading or sulfur wt%, due to the updated calculation for the initial and variable area, accounting for the mass of sulfur directly, as shown in Equations (11), (14), (18). The change in area then influenced the utilisation of the sulfur based on the available reactions sites for further reactions to take place, through the inclusion of area in the precipitation dynamics. The resistance was calculated based on the mass of species within the cell, which decreased with lower sulfur content, as shown in Equations (20) – (22). Experimentally, to capture variations to sulfur loading and wt% simultaneously, both the sulfur mass and electrode thickness needed to be varied together. Due to the calculation for usable electrolyte and sulfur mass within the cathode region, along with diffusion of species between regions based on the concentration gradient, the model results with a change in both thickness and mass together did not align with experimental results; instead, the capacity remains unchanged. Further work is required to investigate and capture this behaviour.

The larger capacity loss Boenke et al. [18] observed experimentally at higher C-rates, hypothesised to be due to transport limitations and degradation mechanisms, was achieved in the model by removing the assumption of total material utilisation, introducing reservoirs and a diffusion mechanism between. Capacity fade unrelated to C-rate was not achieved, as the model did not account for degradation mechanisms. However, a shuttle mechanism was included, to investigate the behaviour with a competing current, as opposed to a loss of shuttled material. When testing shuttle, the Coulombic efficiency reduced and a plateauing charging voltage occurred, aligning with observations by Boenke et al. [18]. The influence of shuttle was minimised at higher C-rates, observed by a spiking in charging voltage towards the end.

## 5 Conclusion

Focusing on the differences between coin and pouch-cells is essential when trying to understand the scale up process from material development to real-world application. This study has investigated what is required within a physics-based model to predict coin-cell behaviour, and the extent to which existing pouch-cells models could be scaled down to achieve this. Whilst maintaining low complexity and minimal assumptions and fitting parameters, this model has improved the capability of understanding and predicting changes when scaling between cell formats. In the process, several features have been introduced to the model to improve the causality of predictions for both cell-formats, and several incorrect assumptions have been removed.

The diffusion mechanism capturing capacity C-rate dependence is only possible due to the removal of the assumption of total electrolyte utilisation, and the inclusion of pore volume parameter used to calculate the excess electrolyte volume and the dissolved sulfur species within, as discussed in Section 7.2. This excess volume varies dependent on the newly introduced scaling parameters, including E/S ratio, cathode thickness and number of layers, along with the updated area calculation. The results shown in Figure 2 confirm that whilst diffusion is one mechanism required to capture C-rate dependence, along with the influence of precipitation dynamics including nucleation, additional mechanisms may be required to more accurately capture coin and pouch-cell capacity variations. Separately, the dissolution bottleneck in coin-cells that results in a temporary loss in active material shows the potential influence degradation mechanisms could have on model predictions, if used to capture a permanent loss of active material. Whilst improvements compared to existing models have been made, further changes are still required for this coin-cell model to be used alongside experimental work. This includes the addition of material and chemistry variations for components within the cell. In addition, detailed information relating to the cell assembly, including component types and manufacturing methods, such as the influence of pressure, also need to be considered when scaling between formats.

## 6 Acknowledgements

The research was completed at Imperial College London and funded as part of the Faraday Institution's LiSTAR (Lithium Sulfur Technology Accelerator Research) Project. The authors acknowledge the significant amount of work completed alongside this by Dr Michael Cornish, who also worked within the LiSTAR project at Imperial College London. In addition, the authors acknowledge the significant amount of additional information provided by Dr. Tom Boenke, the first author of the data used to parameterise the model.

## 7 Appendix: Model Iterations and Discussion

### 7.1 Nomenclature

| Parameter/Variable Name | Description | Units |
|---|---|---|
| $A$ | Avogadro's constant | mol$^{-1}$ |
| $ar$ | Cathode surface area | m$^2$ |
| $ar_{geom}$ | Geometric area | m$^2$ |
| $ar_0$ | Initial cathode surface area | m$^2$ |
| $ar_{growth}$ | Particle growth surface area | m$^2$ |
| $ar_{spec}$ | Specific area | m$^2$g$^{-1}$ |
| $b$ | Fitting parameter (Conductivity) | S m$^2$ mol$^{-1}$ |
| $cath_{mass}$ | Cathode mass | g |
| $C_{S8}, C_{S4}, C_{S1}, C_{Sp(nuc)}, C_{Sp(growth)}$ | Concentration of sulfur species in the cathode | mol L$^{-1}$ |
| $C_{S*}$ | Saturation concentration | mol L$^{-1}$ |
| $C_{cath_{res_{S8}}}, C_{cath_{res_{S4}}}, C_{cath_{res_{S1}}}$ | Concentration of species in the cathode reservoir | mol L$^{-1}$ |
| $C_{sep_{S8}}, C_{sep_{S4}}, C_{sep_{S1}}$ | Concentration of species in the separator | mol L$^{-1}$ |
| $C_{sep_{res_{S8}}}, C_{sep_{res_{S4}}}, C_{sep_{res_{S1}}}$ | Concentration of species in the separator reservoir | mol L$^{-1}$ |
| $C_{Li}$ | Concentration of lithium | mol L$^{-1}$ |
| $C_{Li_0}$ | Initial concentration of lithium | mol L$^{-1}$ |
| $conduc$ | Conductivity | S m$^{-1}$ |
| $conduc_0$ | Initial conductivity | S m$^{-1}$ |
| $D_{Si}$ | Transport coefficient for diffusion (cathode to separator) | s$^{-1}$ |
| $D_{cath_{res_{Sx}}}$ | Transport coefficient for diffusion (cathode to cathode reservoir) | s$^{-1}$ |
| $D_{sep_{res_{Sx}}}$ | Transport coefficient for diffusion (separator to separator reservoir) | s$^{-1}$ |
| $D_{res_{Sx}}$ | Transport coefficient for diffusion (cathode reservoir to separator reservoir) | s$^{-1}$ |
| $\dfrac{dC_{S_x}}{dt}$ | Rate of change of concentration of species, where x = 8, 4, 1 | mol L$^{-1}$ s$^{-1}$ |
| $\dfrac{dC_{sep_{Sx}}}{dt}$ | Rate of change of concentration of species in separator, where x = 8, 4, 1 | mol L$^{-1}$ s$^{-1}$ |
| $\dfrac{dC_{sep_{res_{Sx}}}}{dt}$ | Rate of change of concentration of species in separator reservoir, where x = 8, 4, 1 | mol L$^{-1}$ s$^{-1}$ |
| $\dfrac{dC_{cath_{res_{Sx}}}}{dt}$ | Rate of change of concentration of species in cathode reservoir, where x = 8, 4, 1 | mol L$^{-1}$ s$^{-1}$ |
| $\varepsilon$ | Porosity | |
| $\dfrac{E}{S}$ | Electrolyte to sulfur ratio | |
| $E_{nuc}$ | Nucleated sulfur standard potential | V |
| $E_H^0, E_L^0$ | Standard potentials (high, low) | V |
| F | Faraday's constant | C mol$^{-1}$ |
| $i_H, i_L$ | Exchange current density (high, low) | A m$^{-2}$ |
| $i_H^0, i_L^0$ | Initial exchange current density (high, low) | A m$^{-2}$ |
| I | Applied current | A |
| $k_{nuc}$ | Rate of nucleation | L mol$^{-1}$ m$^{-2}$ s$^{-1}$ |
| ks | Shuttle rate | s$^{-1}$ |
| kp | Precipitation rate | s$^{-1}$ |
| kd | Dissolution rate | s$^{-1}$ |
| l | Cathode thickness | m |
| $M_S$ | Molar mass of sulfur | g mol$^{-1}$ |
| $ns_x$ | Number of S atoms in each species (where x = 8, 4, 1) | |
| $n_H, n_L$ | Number of electrons per reactions | |
| $\rho_{cnt_{pure}}$ | Density of pure BP sheet | g L$^{-1}$ |
| $\rho_{single_{CNT}}$ | Density of single CNT sheets | g L$^{-1}$ |
| $\rho_{sulfur}$ | Density of precipitated sulfur | g L$^{-1}$ |
| r | Radius of sulfur atom | m |
| res | Resistance | Ω |
| R | Gas constant | J mol$^{-1}$ |
| $S_8, S_4, S_1, S_{p\,(nuc)}, S_{p\,(growth)}$ | Mass of sulfur species | g |
| $S_{tot}$ | Total sulfur mass | g |
| T | Temperature | K |
| v | Usable electrolyte volume | L |
| $v_{elec}$ | Electrolyte volume | L |
| $v_{pore\,(cath)}$ | Cathode pore volume | L |
| $v_{tot}$ | Total cathode volume (assuming solid mass) | L |
| $v_{sulfur}$ | Sulfur volume | L |
| $v_{cath\,(CNT)}$ | Cathode (CNT) volume | L |
| $v_{S_p}$ | Volume of precipitated sulfur | L |
| $v_{sep}$ | Volume of electrolyte in the separator region | L |
| $v_{sep_{res}}$ | Volume of electrolyte in the separator reservoir region | L |
| $v_{cath_{res}}$ | Volume of electrolyte in the cathode reservoir region | L |
| V | Cell voltage | V |
| $V_{terminal}$ | Terminal voltage | V |
| $V_{Li_2S}$ | Molar volume of Li$_2$S | m$^3$ mol$^{-1}$ |
| Vr | Reservoir volume | L |
| $x_{cath}$ | Number of cathode layers | |
| $x_{nuc}$ | Number of nucleation layers | |
| γ | Gamma fitting parameter (area) | m$^2$ g$^{-1}$ |
| Y | Fitting parameter | |

*Table 3: Nomenclature*

## 7.2 Upgrade 1: Cathode Porosity and Pore Volume

From analysis of the selection of 0D and 1D models with the literature, most models only consider the influence of cathode porosity, making no clear distinction between porosity and pore volume. Pore volume is the volume within the cathode available for the electrolyte to fill, and for reactions to take place within. Porosity is the ratio between pore volume and the total volume of the cathode. Without clearly defining the relationship between electrolyte and cathode pore volume, most existing models assume the cathode is fully wetted and fully utilised, resulting in model predictions aiming for the maximum theoretical specific capacity, 1,675mAh/g [S] [32]. This is unrealistic; experimental results have found complete sulfur utilisation is rarely achieved [18], [33].

Zhang et al. [12] derived the change in porosity based on the concentration of species, which was calculated using the volume of electrolyte as shown by Equation (3), and the point at which precipitation occurred, based on the saturation concentration. In a real cell, precipitation of sulfur species leads to pore blocking, reducing the pore volume available for reactions to occur within, directly reducing the cathode porosity [10]. To improve the accuracy of model predictions, independent influences on porosity were introduced. The first independent influence was the amount of electrolyte required to fill the cathode pores, which should be calculated based on the cathode pore volume directly. Within the model, the cathode pore volume was referred to as the volume of free space within the cathode at the time of manufacture, and was considered a constant parameter. To calculate this volume, as shown in Equation (8), the model needed to account for the geometric size and thickness of the cathode, the volume of the CNT structure of the cathode, as shown in Equation (5), the volume of sulfur species within the cell, as shown in Equation (6), and the total volume of the cathode, as if it were solid, as shown in Equation (7). These parameters vary when scaling between cell formats; pouch-cells have a reduced E/S ratio, an increased geometric size and a larger number of layers, increasing the likelihood of the electrolyte volume being insufficient to fill the entire cathode pore volume.

The usable pore volume was the volume available for reactions to occur within, and was determined by the relationship between the pore volume and electrolyte volume within the cell. To calculate the usable pore volume, which was also a fixed volume, the distribution of electrolyte across components within the cell needed to be accounted for. Within the proposed model, these components included the cathode and the separator. The model first attempted to fill the cathode pore volume with electrolyte. If there was sufficient electrolyte to do this, the electrolyte then filled the separator pore volume. Any remaining electrolyte created a reservoir of excess electrolyte. This reservoir was split into two regions, the cathode reservoir and separator reservoir, and is discussed in more detail in Section 7.4. The model assumed that the electrolyte must be contained in one of these four regions, and that the regions must be filled in this order. Only the species dissolved in the electrolyte within the cathode contributed to electrochemical reactions. The species were assumed to be homogeneously distributed through the entire electrolyte volume. As previously discussed, the usable pore volume of the cathode, $v$, was calculated as shown in Equations (9) – (14), and influenced the initial area available for reactions at the beginning of cell cycling.

The second independent influence was the amount of precipitated sulfur, which should influence porosity through the blocking of available pore sites. To do this, the porosity equation was updated, as shown by Equations (15) and (16), and calculated based on the relationship between the pore volume and the total cathode volume, derived from the cathode (CNT) volume and precipitated sulfur volume. As species precipitated, the volume of the precipitates increased, reducing the pore volume available for reactions to take place within, and therefore reducing the cathode porosity. Separately, the electrolyte volume influenced the available pore volume, the initial area available for reactions, and the concentration of all species, based on the usable quantity. By separating the dependencies, the precipitation of sulfur became the only direct influence on porosity, allowing the model to more accurately capture the achievable capacity, removing the assumption of total material utilisation.

### 7.2.1 Results

*The equations used within the model to obtain these results are shown in the Appendix (Section 7.6).*

Without accounting for the cathode pore volume, the total electrolyte volume and sulfur mass were assumed to contribute towards electrochemical reactions, and the model aimed to achieve the maximum capacity of 1,675 mAh/g (S) [18], as shown in Figure 3A. When the cathode pore volume was accounted for, the usable volume was calculated dependent on whether the electrolyte volume was in excess, matched or was insufficient to fill the cathode pore volume, as previously discussed and shown in Equations (9) – (14). As the sulfur species were homogeneously distributed throughout the electrolyte, those dissolved within the excess electrolyte did not contribute to electrochemical reactions and were permanently lost, resulting in reduced cell capacity, as shown in Figure 3B.

| Updated Porosity Equation: Equation (16) Zhang et al. [12] Area Equation: Equation (35) ||
|---|---|
| Without accounting for pore volume: *Replace Equations (11) and (14) with $ar0 = ar0$* | Accounting for pore volume: *Equations (9) – (14)* |

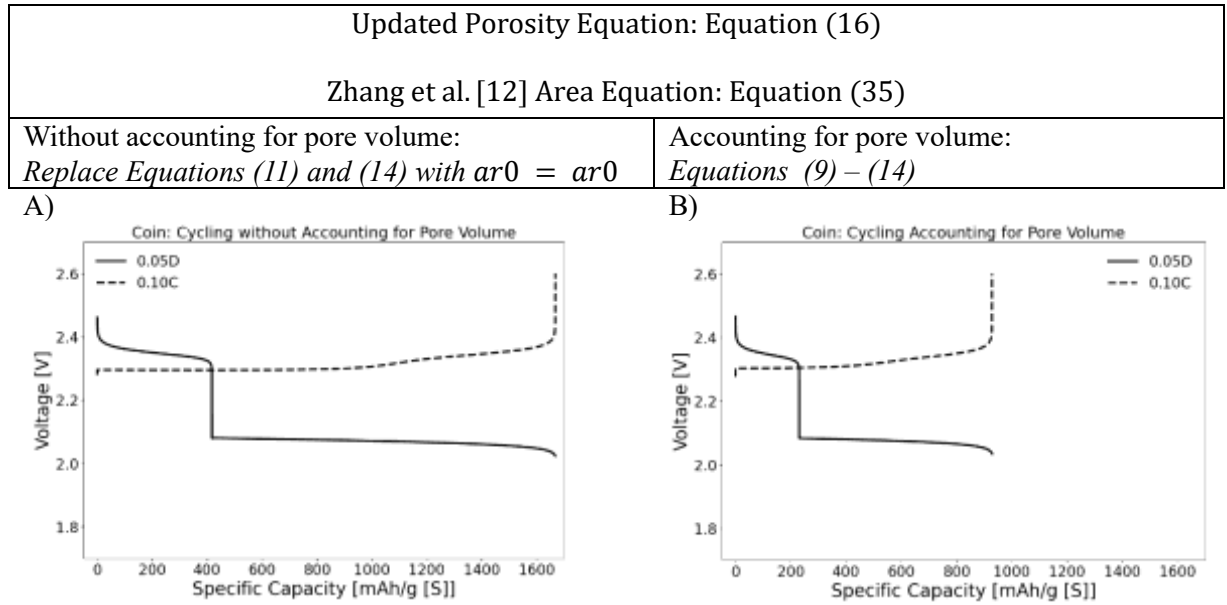

*Figure 3: Model predictions for coin-cell cycling, without vs. with accounting for pore volume.*

Further investigation found that if the electrolyte volume was similar or matched the cathode pore volume, the highest capacity was achieved. This was due to no, or a minimal amount in the excess reservoirs, meaning the total electrolyte volume and dissolved sulfur were contained within the cathode, contributing towards electrochemical reactions and allowing the model to reach maximum capacity. If the electrolyte volume was insufficient to fill all the cathode pores, the cathode was not fully utilised; the portion of the cathode that was filled by electrolyte was fully accessible, with the remaining unfilled portion inaccessible for reactions. This reduced both the usable pore volume and usable sulfur mass, resulting in a lower achieved capacity. This capacity change was irreversible, and although it did not capture degradation directly, it proved the influence of availability of active material on cell performance.

Whilst similar capacities were achieved in both cases when the electrolyte volume did not match the pore volume, if in excess or when insufficient, the causality for the behaviour was different. This was an important consideration for further work into how species dissolved within the reservoir of excess electrolyte changed during cycling, as discussed later in Section 7.4.

### 7.2.2 Overview and Further Work

The addition of pore volume was crucial to accurately predict how the cell utilised the electrolyte volume, sulfur mass and surface area. The differentiation between pore volume and porosity enabled the model to more accurately account for behaviour changes when scaling between cell formats, including capacity changes due to material loss. Coin-cells have a much larger excess of electrolyte in comparison to pouch-cells, meaning when calculating the usable volume this way, a much larger proportion of species was contained outside of the cathode, contributing to a reduction in achievable capacity.

Further testing may be required to understand the extent to which electrolyte influences cell behaviour, beyond the utilisable volume. For example, the influence of excess electrolyte differs at beginning of life where minimal degradation is hypothesised to occur, and long-term cycling where degradation irreversibly consumes active materials, including the electrolyte [34]. In addition, the model did not account for factors such as electrolyte viscosity, solubility of polysulfide species within the electrolyte, or the influence of pressure with varying electrolyte volumes, dependent on cell-format and casing. Further work to understand how these will influence the behaviour of the cell is required, although this may require higher dimensionality to test accurately.

### 7.3 Upgrade 2: Cathode Surface Area and Precipitation Dynamics

Zhang et al. [12] calculated area in relation to the change in porosity during cycling, as derived by Kumaresan et al. [7]. Instead, this model proposed the precipitation of sulfur, leading to pore blocking, directly influenced the available surface area. Separately, it is discussed within the literature that two precipitation mechanisms occur: nucleation and particle growth [10]. It is predicted that precipitation via nucleation, where new nucleation sites are formed for species to precipitate upon, occurs first. Nucleation has a larger influence on the area than particle growth, which occurs afterwards when species precipitate on top of existing nucleation sites [28]. Analysis of the available models shows clear differentiation between precipitation via the nucleation and particle growth mechanisms had rarely been investigated in 0D models, and only within a few 1D models [28], [29]. Models typically assume that precipitation occurs by one mechanism only, which influences area consistently throughout cycling.

#### 7.3.1 Updated Area

To improve the accuracy of the area calculation, and to incorporate the influence of two separate precipitation mechanisms, a new surface area equation was proposed, as shown by Equations (17) and (18). This new equation assumed only precipitation via nucleation influenced area; particle growth precipitated on top of existing nucleation sites, rather than consuming surface area [28]. Similarly to existing models, the initial surface area available at the beginning of each cycle influenced how the area varied throughout the cycle [12]. The initial area was influenced by the usable pore volume, as previously discussed and shown in Equations (11) and (14). To calculate the area, a fitting parameter $\gamma$, as shown in Equation (17), was used, calculated based on the radius of the precipitates and an assumed number of nucleation layers that formed on the surface of the cathode. In addition, a separate particle growth area was introduced, and was calculated based on the concentration of particle growth precipitates, the usable volume of electrolyte and the molar volume of lithium polysulfides. The derivation is included in the Appendix (Section 7.7).

#### 7.3.2 Nucleation and Particle Growth Precipitation Ordinary Differential Equations (ODEs)

To account for two precipitation mechanisms, two separate equations were derived. The nucleation equation, as shown by Equation (30), was derived based on experimental findings by Fan et al. [27], and modelling work by Ren et al. [29] and Danner et al. [28]. Fan et al. [27] stated that for the tested material, nucleation occurred at a critical overpotential of 0.1V, approximate to an absolute cell voltage of 2.05V. This voltage aligned with the transition from high to low plateau in discharge. As this proposed model did not account for chemistry or material variations, this equivalent cell voltage was used and incorporated into an exponential term within the precipitation equation, similarly to the equation used by Ren et al. [29]. The exponential term implied voltage as a driving force; the rate of reaction increased significantly when the cell voltage fell below the nucleation voltage.

Both Ren et al. [29] and Fan et al. [27] found the presence of precipitated species encouraged further precipitation. Similarly to existing models, this was included in the model with the concentration of the corresponding precipitate on the right hand side of the ordinary differential equation (ODE), as shown in Equations (30) and (31). This term also ensured the rate of reaction did not fall below zero when the concentration of precipitates became very small. The surface area was also introduced into both rate equations, creating a circular dependency; as species precipitated via nucleation, the rate of reaction increased, and the available surface area decreased. Once the area was sufficiently small, the rate of nucleation decreased, reducing the rate at which $S_1$ was consumed for nucleation until it fell below the rate at which $S_1$ was being consumed for particle growth precipitation. At this point, the precipitation mechanisms swapped, and particle growth mechanism became dominant. The concentration of $S_1$ was required in both precipitation mechanisms to determine how it was split and ensure the total sulfur mass was conserved throughout cycling.

The particle growth mechanism, as shown by Equation (31), followed the standard form for precipitation used in existing models [8] [10], and was driven by the saturation concentration. The only change was the addition of the area terms. A subtraction term linking the available cathode surface area and the initial cathode surface area, as shown Equation (31), ensured particle growth only occurred during discharge when sufficient nucleation has occurred and reduced the available cathode surface area. When there were no nucleated precipitates, $ar = ar_0$, meaning the rate of particle growth equalled to zero. As nucleated species formed, covering the surface area and creating reaction sites for particle growth to precipitate upon, this term became larger, causing the rate of particle growth to increase. The second area term introduced was the particle growth area, as shown by Equation (19), which was the area created from the formation of particle growth precipitates on top of existing nucleation sites. This term increased the rate of particle growth precipitation as more particle growth precipitates formed. It also ensured nucleation occurred first, followed by particle growth during discharge.

The nucleation, dissolution and precipitation rates varied to determine which reaction was dominant. To satisfy units with the addition of area into the precipitation dynamics, the dissolution and precipitation rate terms became fitted parameters as opposed to physically meaningful rates, each with the unit $\frac{1}{m^2\,s}$. The nucleation rate had units $\frac{L}{mol\,m^2\,s}$.

### 7.3.3 Results

| **Without** area in the precipitation dynamics | **With** area in the precipitation dynamics | |
|---|---|---|
| Porosity: Equation (16) | | |
| Area: Equation (35) | | Area: Equation (18) |
| **Variation 1**: Equation (38) | **Variation 2**: Equation (39) | **Variation 3**: Equations (31) and (30) |

A) B) C)

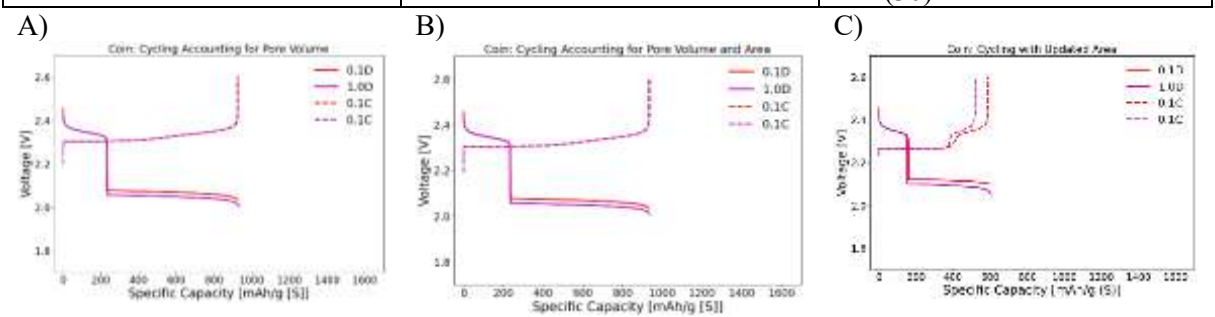

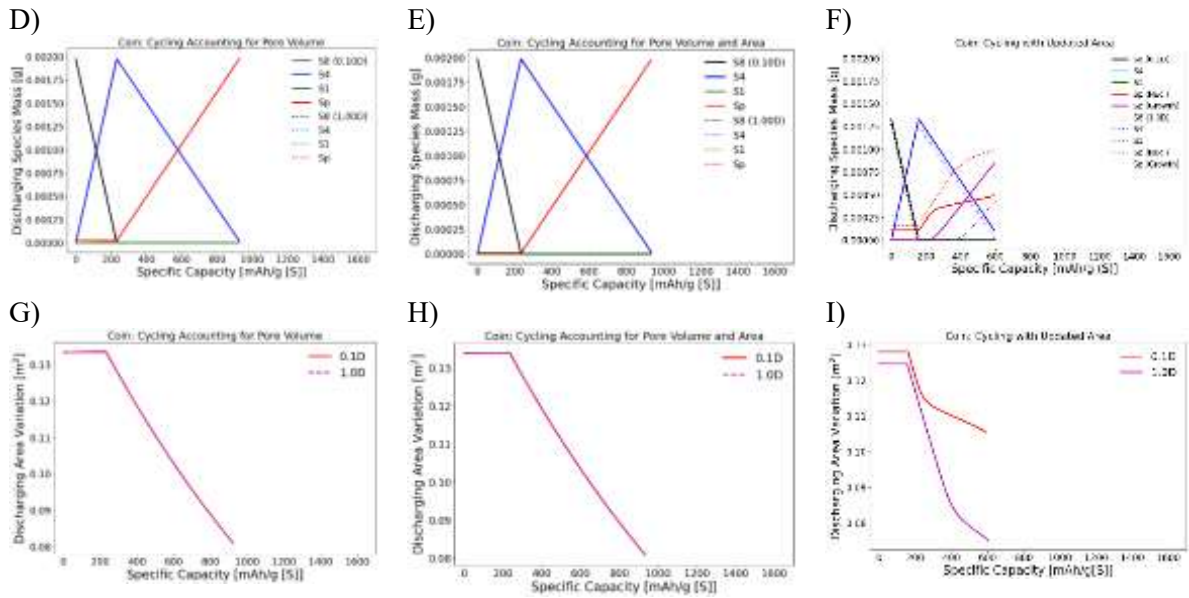

*Figure 4: Model predictions for coin-cell behaviour with the inclusion of area in the precipitation dynamics, and the updated area equation.*

### 7.3.3.1 Variation 1: Zhang Area [12], without Area in Precipitation Dynamics

Variation 1 in Figure 4 shows model predictions using the area equation by Zhang et al. [12]. Despite the updates to differentiate between pore volume and porosity, as discussed in Section 7.2, the model predicts less than half of the cathode surface area was consumed by all precipitating species formed during discharge, as shown in Figure 4G. Although adjustments to the fitting parameters within the area equation could improve the magnitude of predictions, this equation was also unable to capture C-rate variations, as observed in Figure 4A, as the model did not include transport-based changes. Experimentally, the rate of precipitation would be likely to vary with cycling current through the cell based on the rates of reactions, impacting the available area for reactions to occur. Separately, discharge voltage C-rate dependence was observed due to the higher Ohmic change at higher currents. As charging current remains constant, no voltage change during charging was observed.

### 7.3.3.2 Variation 2: Zhang Area [12], with Area in Precipitation Dynamics

The addition of area into the precipitation dynamics created a dependency between the available area and the amount of precipitating species; as precipitation occurred, the available area decreased, which then reduced the rate at which further precipitation would occur. This relationship was essential to include within the model to capture the causality of experimental results. However, as shown in Variation 2 in Figure 4, this addition had minimal impact on behaviour; it was hypothesised this was due the small magnitude of area change in comparison to the larger change in concentration of species.

### 7.3.3.3 Variation 3: Nucleation Area, with Area in Precipitation Dynamics

The update to calculate area based only on precipitation via the nucleation mechanism resulted in a more significant change in behaviour predictions during cycling. As the nucleation mechanism was driven by voltage, the nucleation reaction only began when the voltage dropped below the critical value in discharge, which aligned with the transition from high to low plateau, as shown in Figure 4F. The particle growth reaction was driven by the saturation rate. The competing forces determined the rate at which each reaction took place; when the rate of particle growth exceeded the rate of nucleation, the formation of nucleated precipitates reduced, and the formation of particle growth precipitates increased. The change in area followed the same trend in behaviour; as nucleated precipitates formed, area decreased, and then tended towards a plateau as the rate of nucleation decreased.

When using the updated area equation with increasing C-rate, the initial area at the start of each cycle was lower due to precipitated species not fully dissolving during the previous charge, as shown in Figure 4I. Based on the derivation of the equations, under certain conditions the rate of particle growth could exceed the rate of nucleation before the surface was fully covered. Adjustment to nucleation, dissolution and precipitation rate impact the point at which this occurs, along with the magnitude of area change based on the formation and dissolution of each species. At higher C-rates, the increased driving force for nucleation resulted in more nucleated precipitates forming, and a larger surface area coverage, as shown in Figure 4F, aligning with finding from Ren et al. [29] that the nucleation mechanism is more significant at higher discharge C-rates. This formulation of the area equation resulted in area variations with C-rate.

### 7.3.4 Overview and Further Work

The impact of voltage driven nucleation and saturation concentration driven particle growth highlighted the importance of determining the correct mechanisms, with the correct causality, to obtain experimentally observed behaviour relating to the precipitation dynamics. The inclusion of separate precipitation mechanisms, each occurring at different rates and influencing area separately has enabled the model to capture changes in behaviour with C-rate, including area C-rate dependence. For further development, the ability for the model to account for the influence of pore size and distribution on cell behaviour could be investigated, along with the size and uniformness of precipitates, which vary dependent on C-rate as found by Ren et al. [26].

Although the model captures the correct trends in nucleation and particle growth during discharge, further work is required to correctly capture behaviour during charge. The formulation of the equations meant that an adjustment to the dissolution rate was required to ensure particle growth precipitates dissolved first during charge, before nucleated precipitates. An update to the model would be required to ensure the correct order of reactions under all conditions.

### 7.4 Upgrade 3: Diffusion

The experimental work by Boenke et al. [18] observed both reversible and irreversible capacity changes; reversible changes are linked to C-rate dependence, and irreversible are linked to degradation mechanisms [23]. Whilst irreversible capacity change has been captured in several models through various degradation mechanisms, including [15], [35], reversible capacity C-rate dependence has only been captured in a few 0D models. These include Marinescu et al. [14] which included slow charging to recover capacity loss due to precipitates not fully dissolving, and Cornish et al. [36] which included dynamic diffusion between two 0D models, forming a pseudo-spatial 0D model.

With the addition of pore volume, the model created a reservoir of excess electrolyte when the total volume of electrolyte exceeded the sum of both the cathode and separator pore volume. The species were predicted to be homogeneously distributed throughout the electrolyte and move between the different regions during cycling. Following the methodology of higher order models that capture C-rate dependence through transport mechanisms, along with findings from Cornish et al. [36], a diffusion mechanism to account for the movement of species between regions was introduced. The diffusion was driven by the concentration gradient; species were transported from high to low concentration regions. The cell was split into four regions: the cathode, the separator and the reservoir, comprised of the cathode reservoir and the separator reservoir, as shown in Figure 5. Each of these regions had a different volume and therefore a different concentration of species. Separate rates of diffusion were achieved through separate transport coefficients for diffusion between each region. The movement of species was determined by this rate, the concentration of species in each region, and the ratio of volumes within each region, as shown in Equations (27) - (29), (32) - (34).

With the inclusion of four regions and diffusion between, several assumptions were made. The diffusion mechanism accounted for time only, with no spatial variations; the units of the transport coefficient was

s$^{-1}$. Polysulfide shuttle only occurred to the species within the separator, as this was assumed to be the only region in contact with the anode. Diffusion only occurred between the two neighbouring regions. The reservoir was split into two regions for ease of calculation, but the diffusion rate between the two reservoir regions was set very high, to mimic the behaviour of one larger reservoir. Only the species within the cathode contributed to electrochemical reactions; as the nucleation reaction was driven by voltage, therefore an electrochemical reaction, and must occur before particle growth, the model assumed precipitates only formed in the cathode region. The charged species, $S_8$, $S_4$ and $S_1$ were present in all four regions.

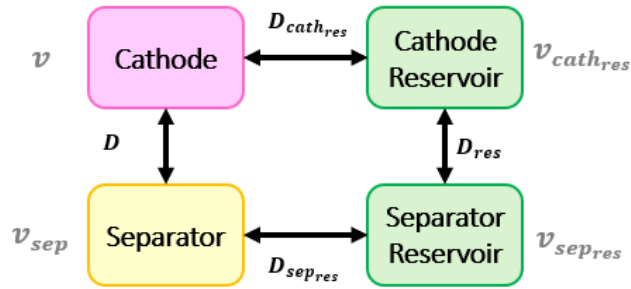

Figure 5: Diagram of Four Reservoir Regions

### 7.4.1 Results

| Without diffusion | With diffusion (constant per species) $D = 4e-3$ | With diffusion (varying per species) $D = 4e-3$ (S8), $4e-4$ (S4), $4e-5$ (S1) |
|---|---|---|

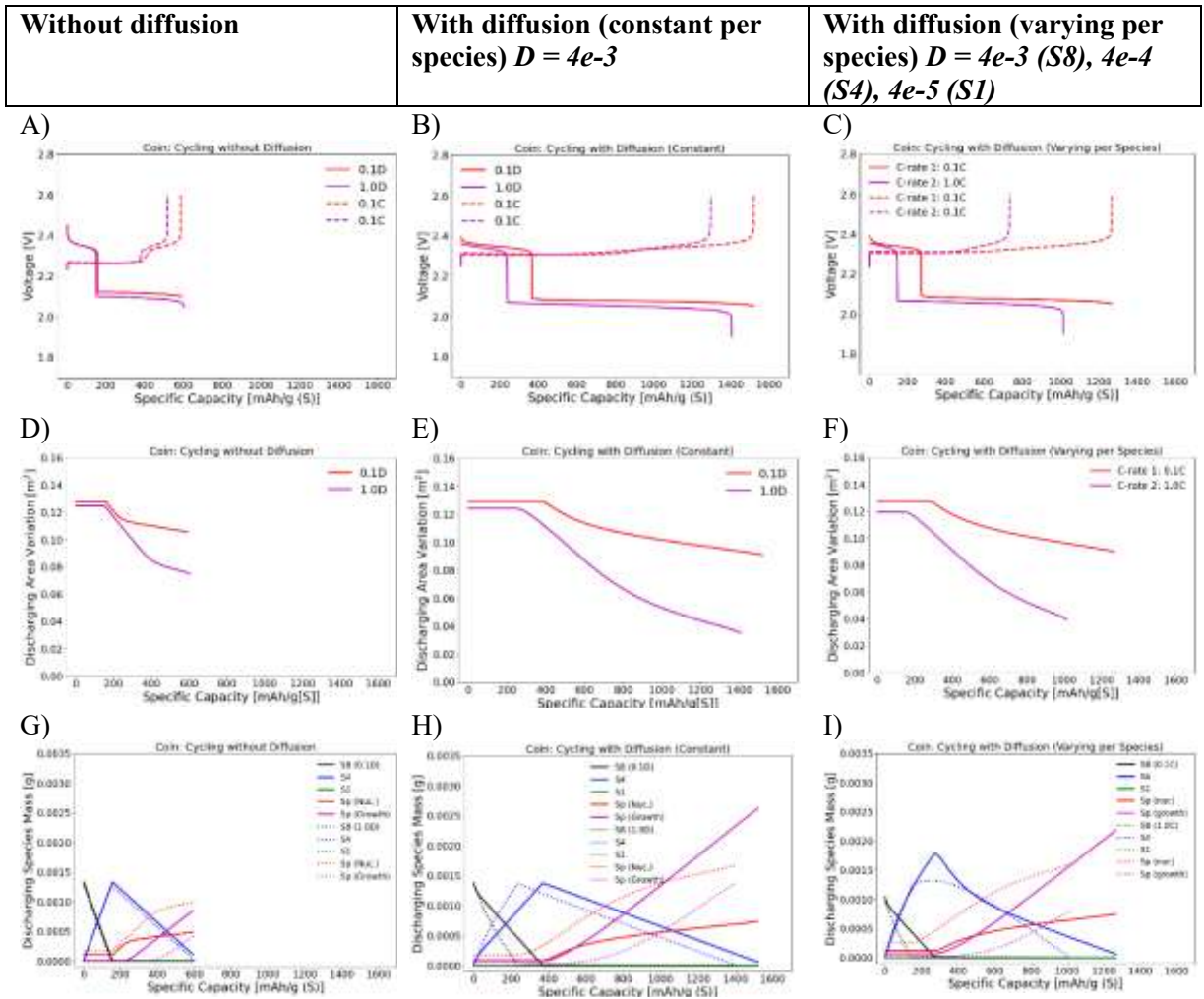

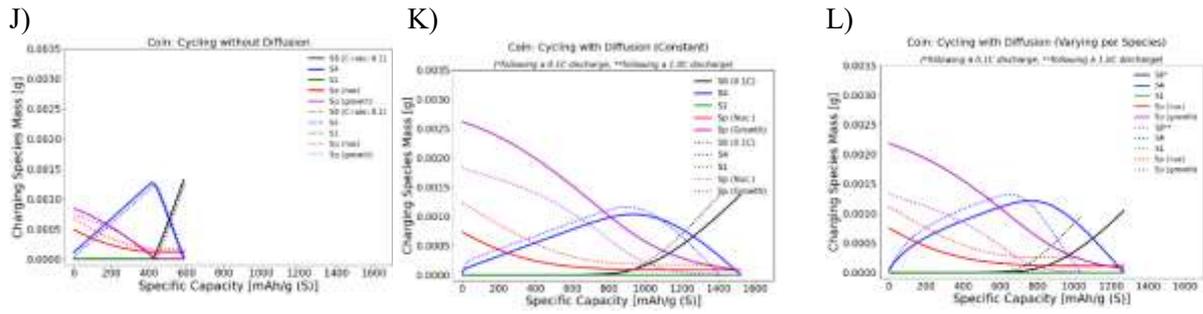

*Figure 6: Model results for coin-cell with and without diffusion, including varying rates of diffusion*

Without diffusion, the model was unable to capture reversible capacity change due to C-rate, as shown in A. The change in capacity observed was due to a loss of active material, and a loss of available surface area, based on the nucleated precipitates not fully dissolving during charge, as shown in Figure 6D, G, J and discussed in Section 7.3.

The model was first tested with the same transport coefficient for each species: $S_8$, $S_4$ and $S_1$. The inclusion of diffusion allowed species to move between region, and be regained into the cathode, increasing the achievable capacity of the cell, as shown in Figure 6B, as a larger mass became available for electrochemical reactions, as shown by Figure 6H compared to G. It also created reversible capacity C-rate dependence, seen by a larger change in capacity at higher C-rates, in both high and low plateaus, in Figure 6B. This was due to faster rates of reaction at higher currents, but a constant rate of diffusion, meaning larger gradients form between regions but insufficient time to diffuse into the cathode, meaning fewer species were available for electrochemical reactions, reducing the achievable capacity. This capacity C-rate dependence occurred in addition to the capacity loss due to inaccessible surface area, corresponding to a larger magnitude of area change, as observed in Figure 6E.

Variations to the transport coefficients per species were then tested; it is understood that the mobility of high-order sulfur species is greater than low-order species [13]. The transport coefficients were assumed constant throughout cycling. As seen from Figure 6C, a lower achievable capacity and a larger C-rate dependence, in both high and low plateau, was observed compared to equal transport coefficients for all species. The slower rate for low-order species resulted in a reduced portion of species being regained from other regions, predominately $S_4$, as shown in Figure 6I compared to H. Further testing confirmed that higher transport coefficients achieved a higher capacity, as more species could be regained from other regions. The magnitude of area change remained similar with varying transport coefficients, as shown in Figure 6F compared to E, although the slower rate resulted in a slightly larger drop in maximum area at the beginning of each discharge, as shown in Figure 6I compared to H.

As previously discussed, there was minimal voltage variation during charging as the C-rate remained constant. However, capacity C-rate dependence was achieved based on the behaviour of the cell during the previous discharge; the final distribution of species at the end of the discharge becomes the initial distribution for the following charge. It is often referred to as the history effect [1], and as shown in Figure 6L, this distribution varies with C-rate.

### 7.4.2 Overview and Further Work

The model changes to incorporate pore volume, electrolyte distribution and the diffusion mechanism were required to capture how the usable species move through the cell during cycling. By creating a reservoir of excess electrolyte, a diffusion mechanism between the four regions within the cell enabled the model to capture higher capacity, along with C-rate dependence. This capacity change due to diffusion was reversible, with higher capacity being re-achieved with the reduction of the current, aligning with experimental findings by Boenke et al. [14].

Further work is required to account for the influence of irreversible changes due to degradation mechanisms, for example, similarly to work by Marinescu et al. [15] who investigated how polysulfide shuttle contributed to irreversible capacity fade due to a permanent loss of shuttled material. The current model only included the shuttle mechanism for the capability to investigate the effect on cycling behaviour and Coulombic efficiency, with no loss of species. For the results shown in this study, the shuttle rate was set to zero. There are various other degradation mechanisms that result in capacity loss, each of which contribute differently to the magnitude and reversibility of the loss. These are also likely to vary dependent on cell format [2]. Degradation to specific components, such as the electrolyte or anode, may also need to be considered to capture the correct behaviour. For example, Boenke et al. [18] predicted varying the charging C-rate would impact degradation to the lithium metal anode due to lithium plating. Further work is required to determine the most significant mechanisms, under which conditions these occur, and the influence on different cell-formats.

## 7.5 Additional Results

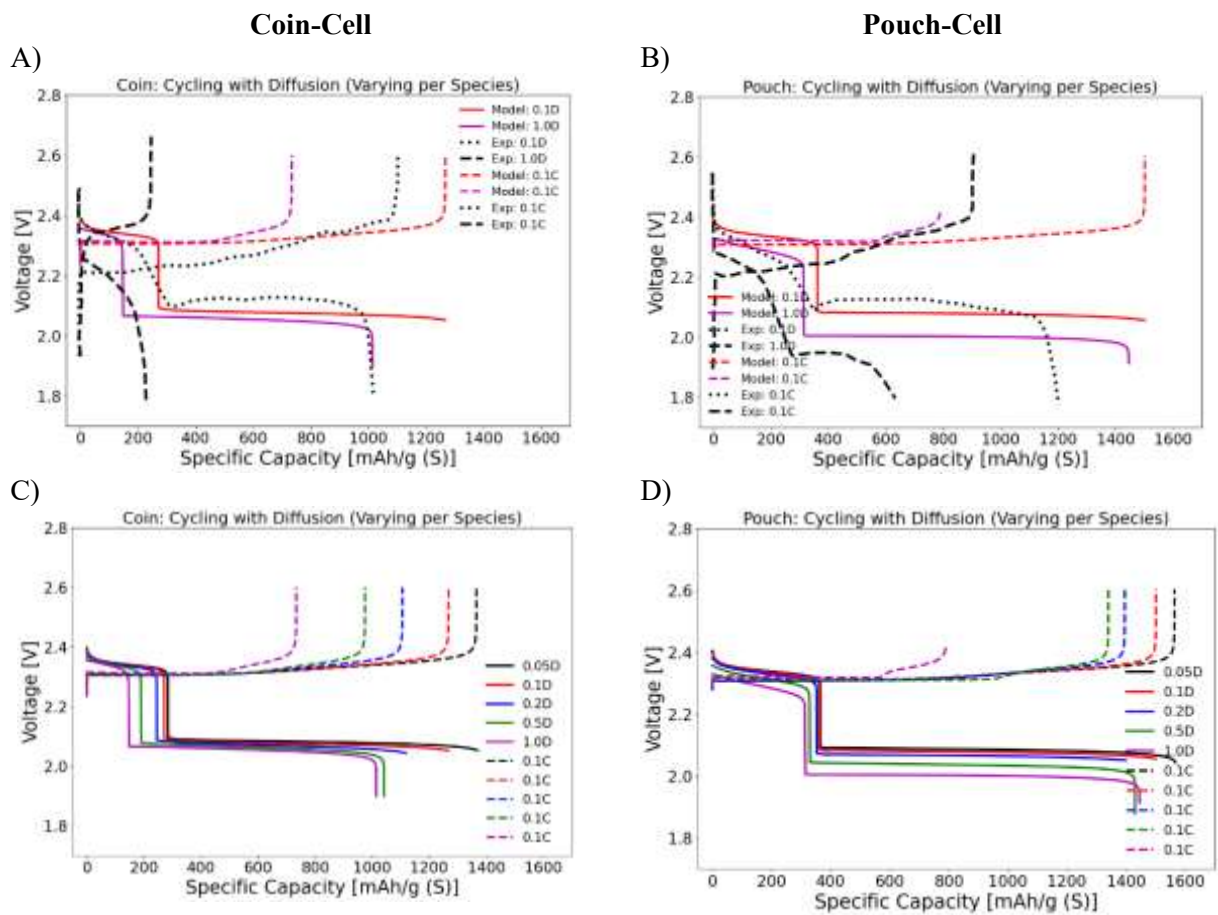

## 7.6 Derivation: Cathode Pore Volume and Porosity

When calculating the influence of pore volume and porosity, the standard area equation, as given by Zhang et al. [12] was used.

$$\text{ar} = \text{ar}_0 \cdot \left(\frac{\varepsilon}{\varepsilon_0}\right)^Y \tag{35}$$

This calculation also only assumed two regions: cathode and reservoir. The diffusion equations were simplified as shown in Equations (36) and (37). v refers to the usable electrolyte volume, and Vr refers to the volume of electrolyte in the reservoir, calculated as the excess electrolyte. The diffusion rate was set to zero to obtain the results shown.

$$\frac{dC_{res_{Sx}}}{dt} = -(D_{Sx} \cdot (C_{res_{Sx}} - C_{Sx})) \tag{36}$$

$$\frac{dC_{Sx}}{dt} = \text{electrochemical} + D_{Sx} \cdot \left((C_{res_{Sx}} - C_{Sx}) \cdot \frac{V_r}{v}\right) \tag{37}$$

Similarly to existing models, only one precipitation mechanism is considered, as shown in Equation (38).

$$\frac{dCS_p}{dt} = \frac{Ms}{\rho} \cdot k_{p/d} \cdot C_{Sp} \cdot (CS^{2-} - CS^*) \tag{38}$$

## 7.7 Derivation: Cathode Surface Area and Precipitation Dynamics Equations

For these model predictions, diffusion rate is also set to zero for all species.

$$\frac{dCS_p}{dt} = \frac{Ms}{\rho} \cdot ar \cdot k_{p/d} \cdot C_{Sp} \cdot (CS^{2-} - CS^*) \tag{39}$$

The particle growth area is calculated by assuming all precipitates on top of the nucleation sites are semi-spheres.

$$vol_{semisphere} = \frac{2}{3}\pi r^3 \tag{40}$$

The volume of Li$_2$S is assumed to match the volume of the hemisphere. The radius is calculated by rearranging this equation.

$$vol_{Li_2S} = vol_{semisphere} \; r^3 \tag{41}$$

$$r = \sqrt[3]{\frac{vol_{Li_2S}}{\frac{2}{3}\pi}} = \left(\frac{3 \cdot vol_{Li_2S}}{2\pi}\right)^{\frac{1}{3}} \tag{42}$$

The radius equation is substituted into the area equation, to calculate the area of the Li$_2$S species.

$$ar_{growth} = 2\pi \left(\left(\frac{3 \cdot vol_{Li_2S}}{2\pi}\right)^{\frac{1}{3}}\right)^2 = 3 \cdot vol_{Li_2S}^{\frac{2}{3}} \tag{43}$$

The volume of Li$_2$S can also be calculated from the volume of Li$_2$S, the concentration of particle growth precipitates and the usable electrolyte volume.

$$vol_{Li_2S} = CS_{p_{growth}} \cdot v \cdot VLi2S \tag{44}$$

This volume is then substituted back into the growth area equation.

$$ar_{growth} = (3 \cdot (CS_{p_{growth}} \cdot v \cdot VLi_2S))^{\frac{2}{3}} \tag{45}$$